\documentclass[onecolumn]{aa}
\usepackage{txfonts}
\usepackage{graphicx}
\begin{document}

\newcommand{\te}{$T_{\rm e}\ $}

\title{Multi-frequency GMRT Observations of the \ion{H}{ii} 
	regions S~201, S~206, and S~209}

\subtitle{Galactic Temperature Gradient} 
\titlerunning{GMRT observations of \ion{H}{ii} regions}

\author{ A. Omar\inst{1}\thanks{aomar@rri.res.in}, 
         J. N. Chengalur\inst{2}\thanks{chengalu@ncra.tifr.res.in} \and 
         D. Anish Roshi\inst{3}\thanks{aroshi@gb.nrao.edu}}
\authorrunning{Omar, Chengalur \& Roshi}

\institute{
       Raman Research Institute, C.V. Raman Avenue, 
       Bangalore 560080, India. 
  \and National Centre for Radio Astrophysics, Post Bag 3, 
       Ganeskhind, Pune 411007, India.
  \and National Radio Astronomical Observatory, Green Bank, 
       West Virginia 24944-0002, USA. }

\date{Received mmddyy/ accepted mmddyy}

\offprints{A. Omar}

\abstract{  We  present  radio  continuum  images  of  three  Galactic
 \ion{H}{ii}  regions, S~201,  S~206,  and S~209  near  232, 327,  and
 610~MHz using  the Giant Meterwave Radio Telescope  (GMRT).  The GMRT
 has a  mix of  short and long  baselines, therefore, even  though the
 data have  high spatial resolution,  the maps are still  sensitive to
 diffuse  extended  emission.   We  find that  all  three  \ion{H}{ii}
 regions have  bright cores surrounded  by diffuse envelopes.   We use
 the high  resolution afforded  by the data  to estimate  the electron
 temperatures  and emission  measures of  the compact  cores  of these
 \ion{H}{ii}  regions.  Our  estimates  of electron  temperatures  are
 consistent  with  a  linear  increase of  electron  temperature  with
 Galacto-centric  distance  for distances  up  to  $\sim 18$~kpc  (the
 distance to the most distant \ion{H}{ii} region in our sample).

}
\maketitle
\section{Introduction}

 A number of studies have  indicated that the electron temperature \te
of  \ion{H}{ii}  region  increases  with  increasing  Galacto-centric
distance (e.g. \cite{deh00} and  references therein). This effect is
attributed  to  a  decrease  in   heavy  elements  abundances  with
Galacto-centric  distance.  A  low   metal  abundance  leads  to  less
effective  cooling  and  consequently  higher  electron  temperature.
These  studies  are  based  either  on estimates  of  \te  from  radio
recombination lines  (RRLs) (which in  turn depend on  corrections for
departures  from   local  thermodynamic  equilibrium   (LTE)  and  for
collisional  broadening  effects),  or   estimates  based  on   line
strengths    of   the   forbidden    line   transitions    of   Oxygen
[\ion{O}{iii}]$\lambda\lambda$4363, 5007 (which are strongly dependent
on  temperature  variations,  if   any,  over  the  observed  volume).
Further,  most   of  these  studies  are  based   on  observations  of
\ion{H}{ii} regions with Galacto-centric distances $R_G \le 15$~kpc
with  very few measurements  of \te  beyond 15~kpc.  Consequently most
determinations  of  metalicities  of  the  outer galaxy  \ion{H}{ii}
regions are based on values of  \te taken from an extrapolation of the
observed   gradient  in   temperature  up   to  about   15~kpc  (e.g.,
\cite{deh00}).   Since the  O/H ratio  (a commonly  used  indicator of
metal  abundance) depends sensitively  on \te,  metalicities of the
outer  galaxy \ion{H}{ii} regions  are poorly  constrained. In  view of
this, it  is important  to get independent  estimates of  the electron
temperatures of \ion{H}{ii} regions in the outer galaxy.

	An independent  measurement of \te can be  obtained from radio
continuum  observations. The ionized  material in  \ion{H}{ii} regions
emits radio continuum through free-free emission.  At sufficiently low
radio frequencies where the nebula is optically thick ($\tau>>1$), the
emergent  radiation  is a  black  body  spectrum,  and therefore,  the
observed brightness  temperature is equal to  the electron temperature
\te. On the other hand,  at sufficiently high radio frequencies, where
the optical depth $\tau$ of  thermal electrons is low ($\tau<<1$), the
observed  brightness is proportional  to the  emission measure  of the
nebula.  Most of the available  radio maps for \ion{H}{ii} regions are
at  high  radio frequencies  (i.e.  above  1.4~GHz,  e.g., Fich  1993,
\cite{balser95}). These maps show  that \ion{H}{ii} regions often have
a bright core with several  knots surrounded by an extended envelop of
diffuse  emission.   These   core--envelope  structures  of  \ion{H}{ii}
regions  imply  that accurate  measurement  of  \te  from low  radio
frequency observations requires  high angular resolution, since, often
only bright compact cores will  be optically thick at frequencies of a
few hundred MHz. This study  presents an analysis of the low-frequency
GMRT  observations  of  three  Galactic  diffuse  \ion{H}{ii}  regions
spanning Galacto-centric distances up to 18~kpc.

The  GMRT  is an  ideal  telescope  for  these observations  since  it
operates at  several low radio  frequency bands, viz., 150,  232, 327,
610, and 1420  MHz and also it has a  hybrid configuration which makes
it  sensitive  to  both  diffuse  emission (on  scales  up  to  $\sim$
45\arcmin~ at 232, 30\arcmin~ at 327, and 17\arcmin~ at 610~MHz) while
also having  the resolution ($\sim$  15\arcsec~ at 232,  10\arcsec ~at
327, and 6\arcsec ~at 610~MHz) to resolve the compact cores.

\section{Observations}

\begin{table*}
\centering
\caption{Observational Details}
\label{tab:obs}
\begin{tabular}{lcccccccccccc}
\hline
\hline
Field Centre &Field Centre &Frequency &Duration of  &Range of  &rms noise in \\
     RA  & Dec & & observation & baselines &the image \\
     (B1950) &(B1950) & (MHz) &(Hours) &(k$\lambda$) &(mJy beam$^{-1}$) \\
\hline
 02$^h$59$^m$12$^s$ & 60\degr 17\arcmin 00\arcsec &231 &8 &0.05--15 & 2.5\\
  & & 616 &8 &0.09--25 & 1.2\\
 03$^h$59$^m$24$^s$ & 51\degr 11\arcmin 00\arcsec & 236 & 9 & 0.05--18 & 7.4\\
  & & 328 & 10 & 0.06--27 & 3.0\\
  & & 613 & 10 & 0.09--49 & 1.2\\
 04$^h$07$^m$18$^s$ & 51\degr 02\arcmin 00\arcsec & 328 & 9  &0.10--26 & 2.0 \\
  & & 613 & 10 & 0.10--50 & 1.0 \\
\hline
\end{tabular}
\end{table*}  

	The observations were carried  out during the period of August
to December,  1999 at three frequency  bands, viz., 232,  327, and 610
MHz.  The GMRT has a  `Y' shaped hybrid configuration of antennas with
six antennas along each of the three arms and twelve antennas randomly
placed in a  compact arrangement near the centre  of `Y' (for details,
see  \cite{swa91}).   The compact  array  at  the  centre is  about  a
kilometer across  and is generally referred as  the ``central square".
Baselines  in  the central  square  (shortest  baseline $\sim  100$~m)
provide sensitivity  to diffuse large scale  emission, while baselines
involving arm  antennas (longest  baseline $\sim 25$~km)  provide high
angular resolution. The GMRT was in its commissioning phase during our
observations, and due to  various debugging and maintenance activities
not all 30 antennas  were available for observations. The observations
were  carried  out with  typically  20  to  25 antennas  in  different
observing sessions.

	The data  were recorded in  the default correlator  mode which
produces visibilities in 128 channels over a user selectable bandwidth
in  multiples of  2  starting from  62.5~kHz  and up  to 16~MHz.   The
observational parameters  are summarized in  table~\ref{tab:obs}.  The
observations  near 610 and  327 MHz  were made  using the  full 16~MHz
bandwidth while observations  near 232 MHz were made  with a bandwidth
of 2~MHz centered at a frequency around which least local interference
has  been  detected in  the  past  observations.   The images  at  all
frequencies are  however made using  data from only one  channel which
corresponds to a bandwidth of 125~kHz at 327 and 610 MHz, and 15.6~kHz
at 232~MHz.  This  restriction was partly because of  a crunch in disk
storage at  the time  when these data  were taken, and  partly because
dynamic range  limitations at the  GMRT at the  time we took  the data
meant that the increase in bandwidth did not result in a proportionate
increase  in sensitivity.   At each  frequency band,  we  observed the
source  for about  8--10  hours, primarily  in  order to  have a  good
$(u,v)$ coverage.

	 For all  the observations, the  source 3C~48 was used  as the
primary flux calibrator.  The flux density of 3C~48  at each frequency
was estimated  using the \cite{baar77} flux densities  of standard VLA
calibrators.  The  phase and amplitude gains of  antennas were derived
from  observations  of  a  secondary  calibrator at  intervals  of  45
minutes. For  observations on  S~206 and S~209,  3C~119 was used  as a
secondary calibrator while 0107+562 was used as a secondary calibrator
for observations on  S~201. Both 3C~119 and 0107+562  are standard VLA
calibrators. The  fluxes of secondary calibrators  were determined via
boot-strapping the fluxes of the primary calibrator 3C~48.

	The  data were  carefully  checked for  interference or  other
problems. At 232  and 327 MHz, a few short baselines  were found to be
corrupted, possibly  by interference, and  were removed.  The  data at
610~MHz were found  to be free from any  interference.  Data reduction
was done in classic AIPS. The calibrated data were Fourier transformed
using appropriate $(u,v)$ ranges, tapers and weights to make different
images, some  of which  are sensitive to  large scale  structures, and
others  which have  the  maximum possible  angular resolution.   These
images were deconvolved using  the `CLEAN' algorithm as implemented in
AIPS task  `IMAGR'. The final gains  of the antennas  were fixed using
several iterations of self-calibration.

   	The variations  in system  temperatures of GMRT  antennas are
currently  not routinely  monitored during  observations.   The system
temperature  at 610  MHz was  measured both  toward the  absolute flux
calibrator 3C~48 and the target source by firing the noise calibration
diodes.  For 327  and 232  MHz images,  the system  temperature toward
3C~48 and target source were obtained using interpolated values of sky
temperature from 408 MHz all--sky map of \cite{haslam82}. A correction
factor equal to the ratio of the system temperature toward the target
source and 3C~48 was applied in the deconvolved image. The deconvolved
images  were  finally  corrected  for the  primary  beam  attenuation,
assuming a Gaussian shape for  the primary beam. The half power points (HPBW)
of the primary  beam of GMRT antenna are estimated  as 1.85, 1.35, and
0.72 degree for 232, 327 and 610 MHz respectively.

\section{Results}

\begin{table*}
\centering
\caption{\ion{H}{ii} Regions -- Results}
\label{tab:res}
\begin{tabular}{lcccccccccc}
\hline
\hline
Name & $\alpha, \delta$~(1950) &Frequency &Flux &Area &\te &$EM$ \\
     &$h$~~~$m$~~~$s$~~~~\degr~~~\arcmin~~~\arcsec &(MHz) &(Jy) &(arcmin$^2$) & (K) &(cm$^{-6}$~pc)\\
\hline
S~201 &02 59 20.1 ~60 16 10 &231 &0.78 &16 &$7070\pm1100$ 
&1.02($\pm0.05$)$~\times~10^{5}$ \\
 & &616 &1.15 &38 & & \\
S~206 &03 59 24.0 ~51 11 00 &236 &16.3 &329 &$8350\pm1600$ 
&3.93($\pm0.40$)$~\times~10^{5}$\\
 & &328 &18.2 &347 & & \\
 & &613 &20.0 &350 & & \\
S~209 &04 07 20.1 ~51 02 30 &236 &13.4 &267 &$10855\pm3670$ 
&2.58($\pm0.29$)$~\times~10^{5}$\\
 & &328 &16.6 &372 & & \\
 & &613 &17.0 &386 & & \\
\hline
\end{tabular}
\end{table*}

\begin{figure*}
\resizebox{6cm}{!}{\includegraphics[angle=-90]{fig1.ps}}
\resizebox{6cm}{!}{\includegraphics[angle=-90]{fig2.ps}}
\resizebox{6cm}{!}{\includegraphics[angle=-90]{fig3.ps}}
\vskip 0.1cm
\resizebox{6cm}{!}{\includegraphics[angle=-90]{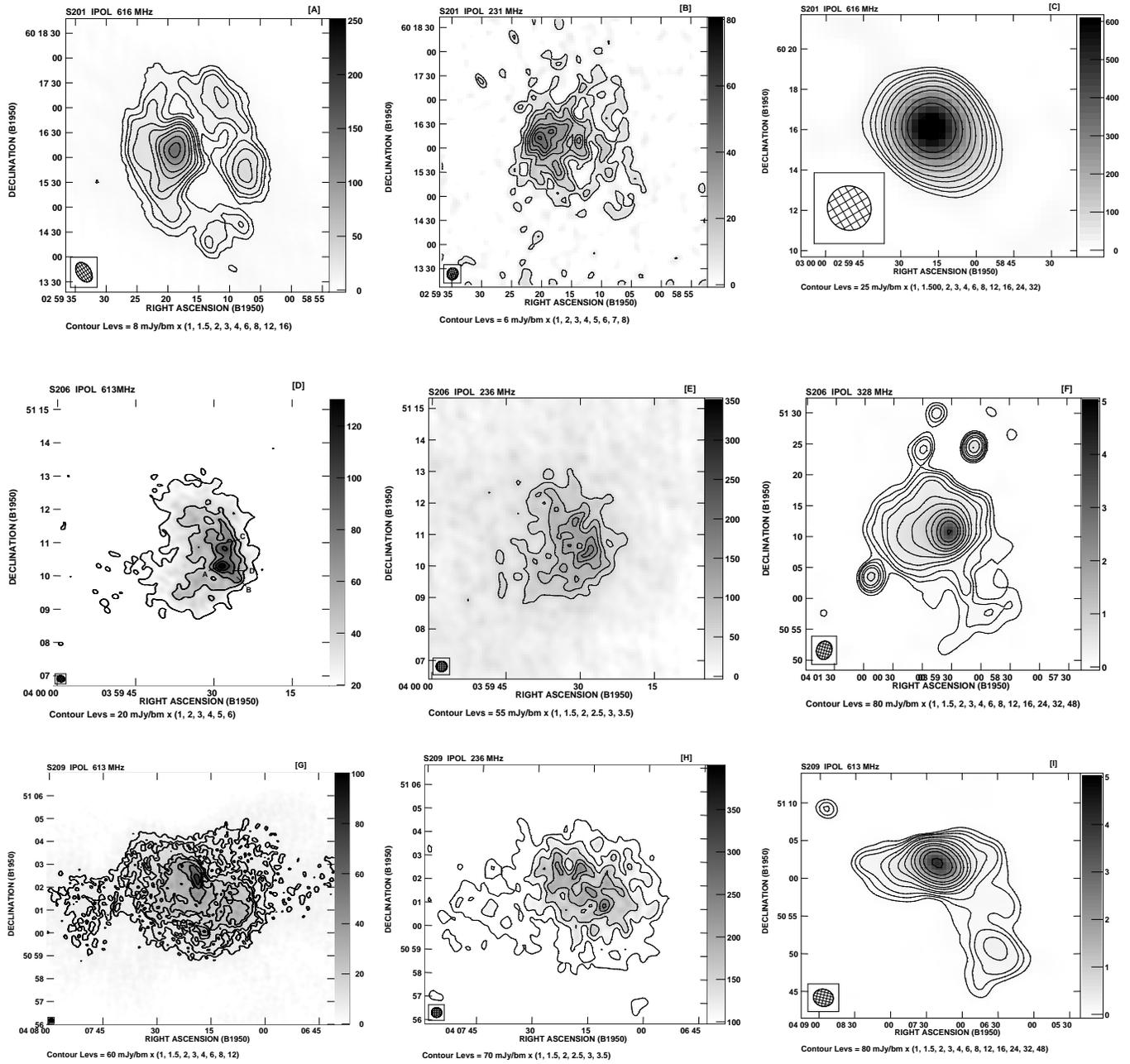}}
\resizebox{6cm}{!}{\includegraphics[angle=-90]{fig5.ps}}
\resizebox{6cm}{!}{\includegraphics[angle=-90]{fig6.ps}}
\resizebox{6cm}{!}{\includegraphics[angle=-90]{fig7.ps}}
\resizebox{6cm}{!}{\includegraphics[angle=-90]{fig8.ps}}
\resizebox{6cm}{!}{\includegraphics[angle=-90]{fig9.ps}}
\caption{
{\bf [A]} S~201 at 616~MHz. The angular resolution is 
          $26\arcsec{~\times~}17\arcsec$. 
{\bf [B]} S~201 at 231~MHz. The angular resolution is
          $15\arcsec{~\times~}13\arcsec$. 
{\bf [C]} S~201 at 616~MHz made using u-v range 0--1 k$\lambda$ only. The angular resolution is 
$133\arcsec{\times}129\arcsec$. 
{\bf [D]} S~206 at 613 MHz. The angular resolution is 
          $13\arcsec{~\times~}11\arcsec$. The regions marked as A, B, C, \& D are from 
the 5 GHz image of Deharveng et al. (1976).  
{\bf [E]} S~206 at 236~MHz. The angular resolution is $20\arcsec{~\times~}20\arcsec$. 
{\bf [F]} 328 MHz image of S~206 made using u-v range 0--1 k$\lambda$ only.
          The synthesized beam is 180\arcsec $\times$ 149\arcsec. 
{\bf [G]} S~209 at 613 MHz. The angular resolution is $10\arcsec{~\times~}10\arcsec$.
{\bf [H]} S~209 at 236 MHz. The angular resolution is $25\arcsec{~\times~}25\arcsec$. 
{\bf [I]} 613 MHz  image of S~209 made using u-v range only up to 
          1~k$\lambda$. The angular resolution is 160\arcsec $\times$
          136\arcsec. }
\label{fig:maps}
\end{figure*}

	Images of  S~201 are shown in  Fig.~\ref{fig:maps}[A], [B], \&
[C].   S~201, ($l=138.48,  b=1.64$  ; also  known  as IC  1848), at  a
Galacto-centric  distance of  $10.5  \pm  1$ kpc,  is  believed to  be
excited due  to a  single star of  spectral type  O9.5 (\cite{mam89}).
High resolution ($\sim$5\arcsec)  15~GHz radio continuum image reveals
a bright arc like core with multiple peaks of emission (\cite{FHC87}).
The 616~MHz GMRT low resolution image (Fig.~\ref{fig:maps}~[C]) traces
diffuse emission  extending up to $\sim$5\arcmin~  which is consistent
with  the 1.4~GHz  VLA image  of \cite{fich93}.   The  high resolution
231~MHz  GMRT image (Fig.~\ref{fig:maps}[B])  shows the  core to  be a
complex  structure consisting of  several unresolved  compact sources.
The diffuse  nebulosity extending toward the  west of the  core in the
231 MHz  image (Fig.~\ref{fig:maps}[B]) is consistent  with the 15~GHz
radio image of Felli et al. (1987).

	Images of  S~206 are shown in  Fig.~\ref{fig:maps}[D], [E], \&
[F].  S~206,  ($l=150.74$, $b=-0.75$; also  known as NGC~1491),  is an
evolved \ion{H}{ii}  region at a Galacto-centric  distance of 11.1~kpc
(\cite{deh00}). The excitation is believed  to be provided by a single
O5  star (\cite{cram74}).   The 5~GHz  radio continuum  image  shows a
classic blister type  morphology (Fig.~4 in Deharveng et  al. 1976) as
described  in \cite{icke80}.   Our high  resolution images  at 613~MHz
(Fig.~\ref{fig:maps}[D]), 236~MHz  (Fig.~\ref{fig:maps}[E]) as well as
328~MHz (not  shown) show  good correspondence to  the 5~GHz  image of
\cite{deh76}.       The     low      resolution      328~MHz     image
(Fig.~\ref{fig:maps}[F])   shows  a   large  low   intensity  envelope
surrounding the core emission.

	Images  of S~209 are  shown in  Fig.~\ref{fig:maps}[G], [H],\&
[I].  S~209 ($l=151.6$, $b=-0.24$; also  known as RAFGL~550) is one of
the most distant ($R_G = 17.7$~kpc, \cite{deh00}) Galactic \ion{H}{ii}
region. Although  the mean size of \ion{H}{ii}  regions decreases with
increasing Galacto-centric distance  (\cite{fich84}), S~209 is unusual
in  that   it  has   a  very  large   size  ($\sim  50$~pc)   for  its
Galacto-centric distance.  The excitation is provided by  a cluster of
OB   stars  (\cite{chini84}).    Our  high   resolution   613~MHz  map
(Fig.~\ref{fig:maps}[G])  shows  the core  region  to  consist of  an
asymmetric, incomplete ring like  structure. The high resolution image
at 328~MHz  (not shown  here) is morphologically  very similar  to the
613~MHz  image.   The  236~MHz  image  (Fig.~\ref{fig:maps}[H])  while
showing overall similarity to the  613~MHz and 328~MHz maps, does show
some difference in the core region.   We are unsure why this should be
so.  The  low resolution 613~MHz  image (Fig.~\ref{fig:maps}[I]) shows
that this  region too has  an extremely large, low  surface brightness
envelope, which has also been seen at 2.7 GHz image of \cite{wal75}.

\section{Discussion}

\begin{figure*}
\resizebox{6cm}{!}{\includegraphics[angle=0]{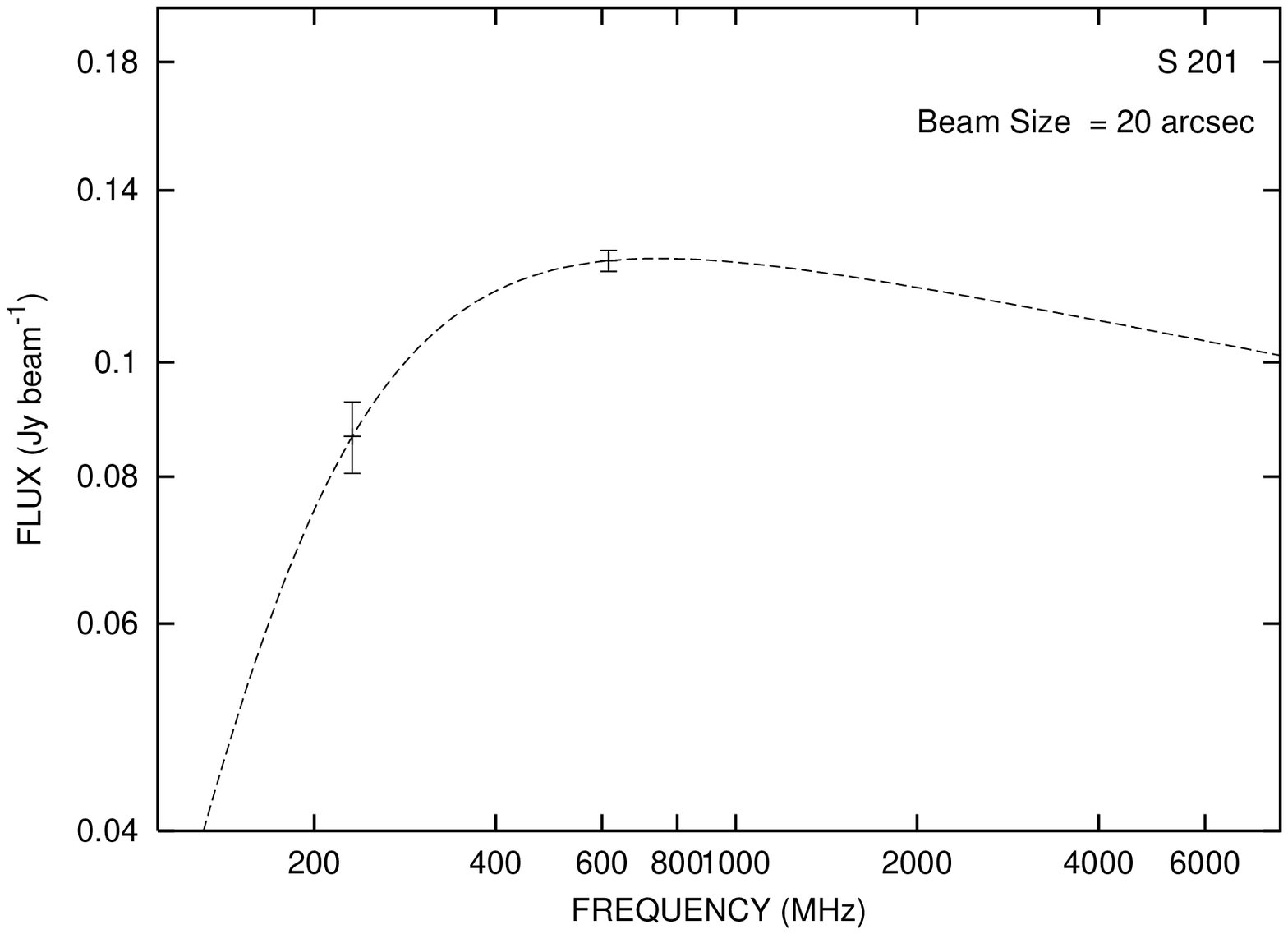}}
\resizebox{6cm}{!}{\includegraphics[angle=0]{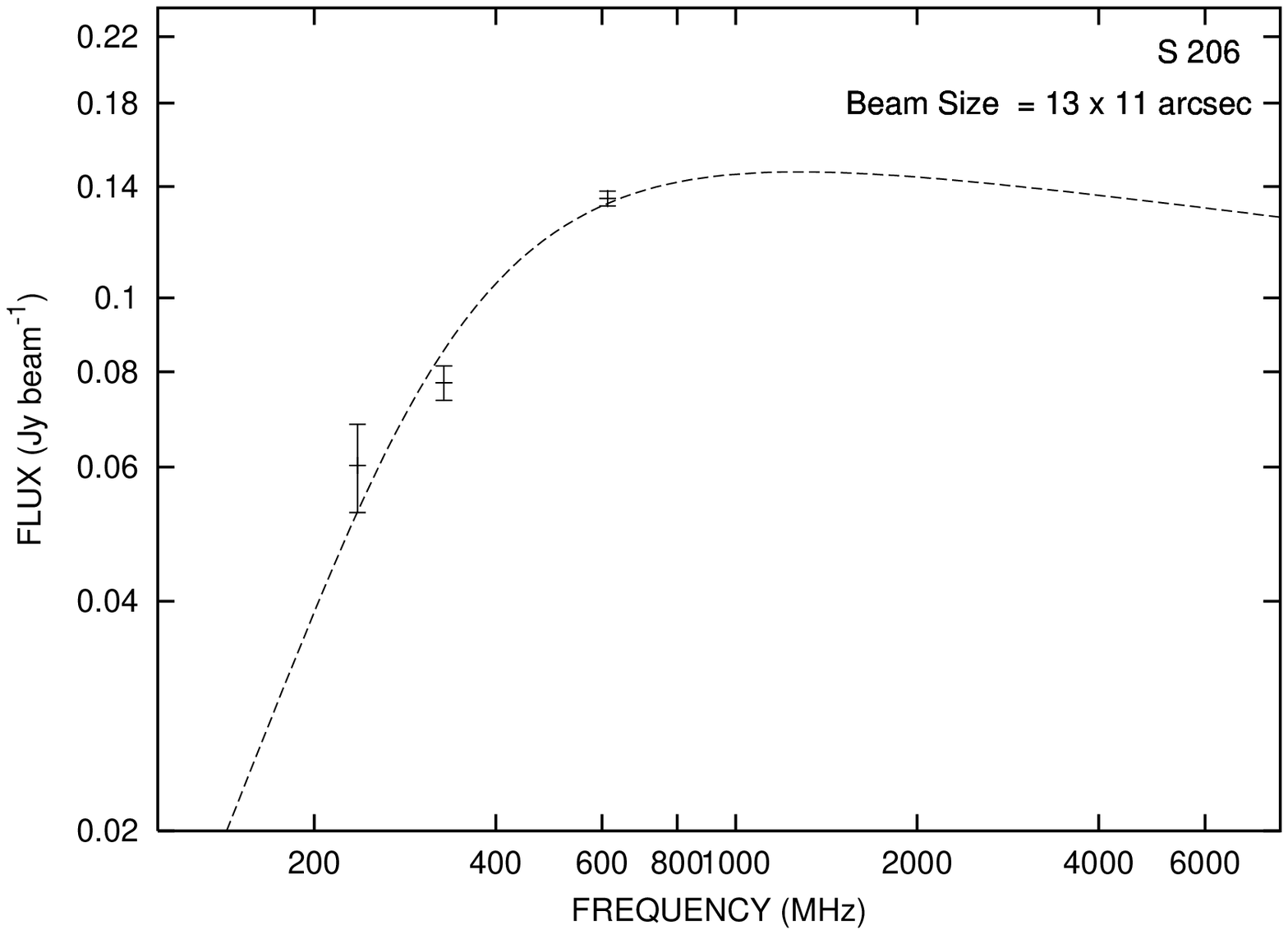}}
\resizebox{6cm}{!}{\includegraphics[angle=0]{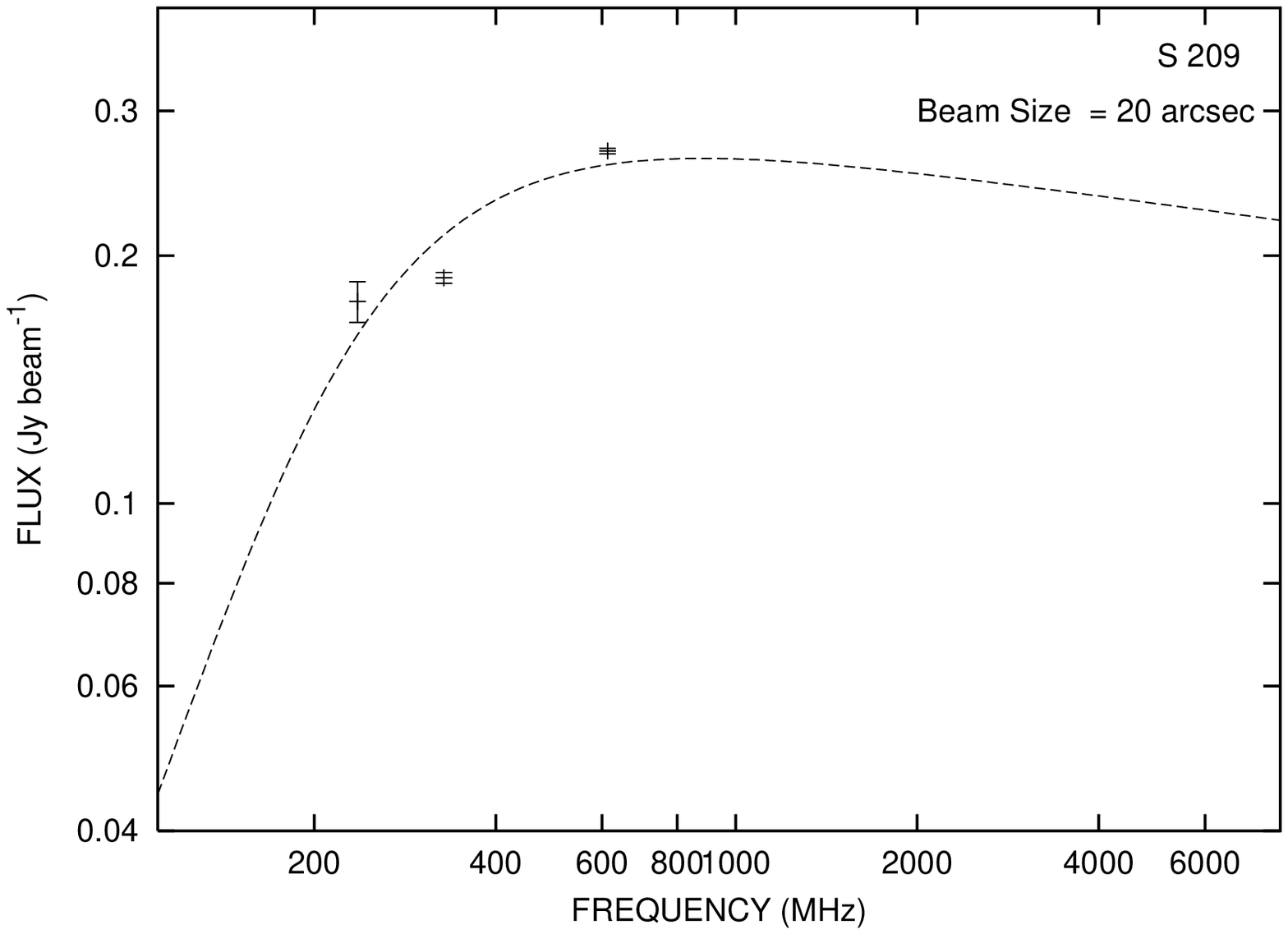}}
\caption{ Model and observed fluxes for the cores of the \ion{H}{II}
          regions observed at the GMRT. The estimated emission measures
          and electron temperatures are listed in table~\ref{tab:res}. }
\label{fig:spec}
\end{figure*}

	We  use  these  low  frequency  images  to  estimate  electron
temperatures and  emission measures  of the compact  cores of  the 
\ion{H}{ii}
regions.  If  we  approximate   these  cores  to  be  homogeneous  and
spherically symmetric, then the flux $S$ is given by

\begin{equation}
\label{eqn:flux} 
S = 3.07 \times 10^{-2} T_\mathrm{e} \nu^2 \Omega (1-\mathrm{e}^{-\tau(\nu)})
\end{equation}
\begin{equation}
\tau(\nu) = 1.643 a \times 10^5 \nu^{-2.1} EM\ T_\mathrm{e}^{-1.35}
\end{equation}

\noindent (\cite{mez67})  where $S$ is the integrated  flux density in
Jy, \te is the electron  temperature in Kelvin, $\nu$ is the frequency
of observation  in MHz, $\tau$ is  the optical depth,  $\Omega$ is the
solid angle subtended by the source in steradian, (which in this case,
since the  cores are unresolved, is  taken to be  the synthesized beam
size), and $EM$ is the  emission measure in cm$^{-6}$~pc. The emission
measure $EM$  is defined  as $\int n_\mathrm{e}{^2}  \mathrm{d}l$; the
integral being  taken along  the line of  sight and averaged  over the
beam.  $a$  is  a  correction  factor  which  depends  both  upon  the
temperature and  frequency. We  have used an  average value of  $a$ as
0.98 (using table~6 of  \cite{mez67}) for the frequency range 200--600
MHz and  \te $\sim$  10\,000~K.  The \ion{H}{ii}  region cores  can be
modeled  by solving equations  (1) and  (2) iteratively  for different
$EM$  and   \te.   The   fitting  procedure  converges   rapidly  when
observations  at   least  two   frequencies  are  available   and  the
frequencies are such that the \ion{H}{ii} region is optically thick at
one frequency and optically thin at the other.

	We measured the peak  flux densities of cores after convolving
the  images of  a \ion{H}{ii}  region  at different  frequencies to  a
common angular resolution (i.e. the  source size $\Omega$ was taken to
be  $1.133~\times  \theta_a  \times  \theta_b$, where  $\theta_a$  and
$\theta_b$ are  the half power  points of the common  convolved beam).
The  best fit  values for  \te  and $EM$  as obtained  from the  fitting
procedure described  above are listed in  table~\ref{tab:res}, and the
observed  and model  fluxes  are plotted  in Fig.~\ref{fig:spec}.  The
columns in table~\ref{tab:res} are  as follows.  col.~(1): Name of the
\ion{H}{ii}   region,    col.~(2):   Coordinates   (right   ascension,
declination)  of the core  for which  the electron temperature  has been
measured, col.~(3): The frequency of observation, col.~(4): Integrated
flux of the entire \ion{H}{ii} region, col.~(5): Area over which radio
emission is detected,  and over which the flux  has been integrated to
get  the  value  listed  in  col.~(4),  col.~(6):  Estimated  electron
temperature of  the core, col.~(7): Estimated emission  measure of the
core.

The electron  temperature of S~201 is estimated  to be $7070\pm1100$~K
toward the peak  radio emission.  The earlier estimate  for \te toward
S~201     was    $\sim5000$~K     based     on    non-detection     of
[\ion{O}{iii}]$\lambda\lambda$4959, 5007 (\cite{mam89}).  The electron
temperature  of $8350  \pm  1600$~K,  derived for  the  core of  S~206
(knot--A  in Fig.~\ref{fig:maps}[C]) is  in reasonable  agreement with
previous  measurements,  viz.   $8400\pm  800$~K  obtained  using  the
H$94\alpha$ recombination line  by \cite{car81}, and $9118$~K obtained
from   the   [\ion{O}{iii}]$\lambda\lambda$4363,   5007  lines   ratio
(\cite{deh00}).  The emission  measure  is $3.93(\pm0.40)~\times~10^5$
cm$^{-6}$~pc, consistent with the  value obtained by \cite{deh76}. For
S~209,  the  electron  temperature  corresponding to  the  peak  radio
emission at 613  MHz is estimated to be  $10\,855\pm 3670$~K, somewhat
higher than the value of $8280$~K obtained using the H$137\beta$ 
recombination line by \cite{chu78} but  in reasonable agreement
with the estimate of $11\,000$~K which was derived from H$91\alpha$ \&
H$114\beta$ recombination lines (\cite{balser94}).

\begin{figure}
\resizebox{8cm}{!}{\includegraphics[angle=0]{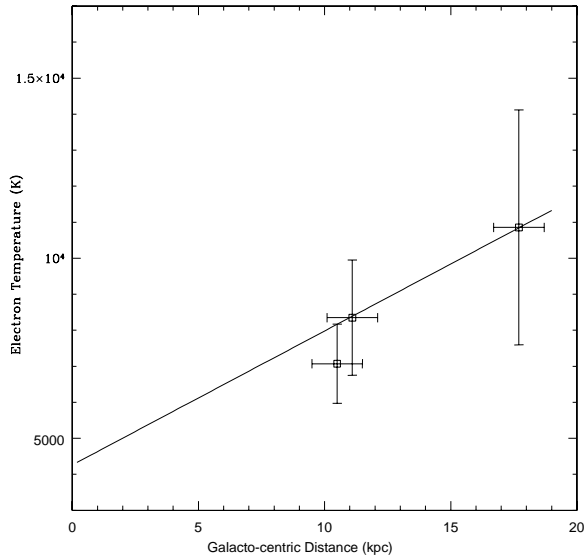}}
\caption{ The electron temperature vs. Galacto-centric distance for three
 \ion{H}{ii} regions studied in this paper. The solid line is the 
relationship
 derived by Deharveng et al. (2000) based on a sample of six \ion{H}{ii} 
regions
 spanning a Galacto-centric distance from 6.6 to 14.8~kpc. 
}
\label{fig:Te}
\end{figure}

Figure~\ref{fig:Te}  is  a  plot   of  the  electron  temperature  vs.
Galacto-centric distance for the  three \ion{H}{ii} regions studied in
this paper. The  solid line is the relationship  obtained by Deharveng
et  al. (2000)  from  a  sample of  six  \ion{H}{ii} regions  spanning
Galacto-centric distances from  6.6 to 14.8~kpc.  The data  for S~209 shows
that this relationship appears to be valid even out to Galacto-centric
distance of $\sim 18$~kpc.

 If there are systematic radial temperature gradients within the cores
of \ion{H}{ii}  regions, the \te  estimated by various  methods, viz.,
radio  continuum, RRLs,  and  [\ion{O}{iii}]$\lambda\lambda$4363, 5007
lines ratio  (all of which  probe different physical regions)  would be
discordant. The radio continuum  observations at low frequencies
are more sensitive to outermost regions of the core of the nebula (due
to  the high  optical depths  at these  frequencies).  The temperature
estimates  from high  frequency  RRLs (which is  where  most of  the
observations   exist)   are    weighted   toward   regions   of   low
temperature. Finally,  estimates of  \te from the  [\ion{O}{iii}] lines
ratio are expected to be  weighted toward high temperature regions due
to  the  high cooling  rate  provided  by  [\ion{O}{iii}] lines. Since,  our
estimates of  \te are in  general consistent with those  obtained from
RRLs as  well as from [\ion{O}{iii}] lines,  any temperature gradients
within the  {\it cores} of  these \ion{H}{ii} regions must  be smaller
than   the    combined   uncertainties   in    these   different   \te
measurements.  Similar concordance  between \te  measured  using these
different methods have been obtained for W~51 (\cite{ravi95}) and M~17
(\cite{ravi96}, \cite{wilson97}).  Several  authors have discussed the
possibility of  small scale temperature  fluctuations in the  cores of
\ion{H}{ii}  regions (\cite{peimbert67}, \cite{rubin98} and
references  therein),   these  cannot  be  ruled  out   based  on  our
observations alone).

\section{Conclusions}

Three outer  galaxy \ion{H}{ii} regions,  S~201, S~206 and  S~209 have
been imaged at meter wavelengths  using the GMRT.  The images of these
\ion{H}{ii} regions have been obtained  at a resolution of less than a
pc. This is the highest resolution achieved for any \ion{H}{ii} region
at such low radio frequencies.   All three \ion{H}{ii} regions show structures down
to our resolution  limit.  The high resolution images  near 610 MHz of
these \ion{H}{ii}  regions show a  good correspondence with  the radio
continuum images  at $cm$ wavelengths.  The low  resolution radio images
show that  these \ion{H}{ii} regions  are surrounded by  large diffuse
envelopes.  The  high resolution radio  images have allowed us  to get
estimates of \te of these \ion{H}{ii} regions. From these measurements
we find that : \\

\noindent (1) the estimates of \te are in general consistent with that
obtained from  RRLs and [\ion{O}{iii}]$\lambda\lambda$4363,  5007 line
measurements, and \\
\noindent (2)  the measured temperatures are consistent  with a linear
increase  of  \te  with   Galacto-centric  distance  until  $R_G  \sim
18$~kpc.

\begin{acknowledgements}

We  thank  the  staff  of   the  GMRT  that  made  these  observations
possible.  The  GMRT   is  run  by  the  National   Centre  for  Radio
Astrophysics of  the Tata Institute  of Fundamental Research.   We are
grateful to K.S.  Dwarakanath and R. N.  Mohan for useful discussions.
We  thank (late)  K.   R. Anantharamaiah  for  useful suggestions  and
comments on an earlier version of this paper.  We are also grateful to
one of  the referees, L.   Deharveng, for many comments  which greatly
improved  the  contents  and  readability  of this  paper.  AO  thanks
NCRA-TIFR for providing hospitality during his stay at NCRA.

\end{acknowledgements}

\end{document}